\documentclass[%
 reprint,
 amsmath,amssymb,
 aps,
 pre,
]{revtex4-2}

\usepackage{graphicx}
\usepackage{dcolumn}
\usepackage{bm}
\usepackage{amssymb}
\usepackage{float}
\usepackage{color}
\usepackage{hyperref}
\usepackage{subfigure}
\usepackage{braket}
\usepackage{cases}

\newcommand{\fn}[2]{\mathinner{#1\mathopen{\left(#2\right)}}}
\newcommand{\spD}[1]{\fn{\tilde{\chi}_{_V}}{#1}}

\begin{document}


\title{\textcolor{black}{Hyperuniformity scaling of maximally random jammed packings of two-dimensional binary disks}}

\author{Charles Emmett Maher}
 \affiliation{Department of Chemistry, Princeton University, New Jersey 08544, USA}
\author{Salvatore Torquato}%
 \email{torquato@princeton.edu}
\affiliation{Department of Chemistry, Princeton University, Princeton, New Jersey 08544, USA}
\affiliation{Department of Physics, Princeton University, Princeton, New Jersey 08544, USA}
\affiliation{Princeton Institute for the Science and Technology of Materials, Princeton University, Princeton, New Jersey 08544, USA}
\affiliation{Program in Applied and Computational Mathematics, Princeton University, Princeton, New Jersey 08544, USA}

\date{\today}

\begin{abstract}
Jammed (mechanically rigid) polydisperse circular-disk packings in two dimensions (2D) are popular models for structural glass formers. 
Maximally random jammed (MRJ) states, which are the most disordered packings subject to strict jamming, have been shown to be hyperuniform.
The characterization of the hyperuniformity of MRJ circular-disk packings has covered only a very small part of the possible parameter space for the disk-size distributions.
Hyperuniform heterogeneous media are those that anomalously suppress large-scale volume-fraction fluctuations compared to those in typical disordered systems, i.e., their spectral densities $\Tilde{\chi}_{_V}(\mathbf{k})$ tend to zero as the wavenumber $k\equiv|\mathbf{k}|$ tends to zero and are often described by the power-law $\Tilde{\chi}_{_V}(\mathbf{k})\sim k^{\alpha}$ as $k\rightarrow0$ where $\alpha$ is the so-called hyperuniformity scaling exponent.
In this work, we generate and characterize the structure of strictly jammed binary circular-disk packings with a size ratio $\beta = D_{L}/D_S$, where $D_L$ and $D_S$ are the large and small disk diameters, respectively, and molar ratio of the two disk sizes is 1:1.
In particular, by characterizing the rattler fraction $\phi_R$, the fraction of configurations in an ensemble with fixed $\beta$ that are isostatic, and the $n$-fold orientational order metrics $\psi_n$ of ensembles of packings with a wide range of size ratios $\beta$, we show that size ratios $1.2\lesssim \beta\lesssim\textcolor{black}{2.0}$ produce maximally random jammed (MRJ)-like states, which we show are the most disordered strictly jammed packings according to several criteria.
Using the large-$R$ scaling of the volume fraction variance $\sigma_{_V}^2(R)$ associated with a spherical sampling window of radius $R$, we extract the hyperuniformity scaling exponent $\alpha$ from these packings, and find \textcolor{black}{the function $\alpha(\beta)$ is maximized at $\beta$ = 1.4 (with $\alpha = 0.450\pm0.002$) within the range $1.2\leq\beta\leq2.0$. Just outside of this range of $\beta$ values, $\alpha(\beta)$ begins to decrease more quickly, and far outside of this range the packings are nonhyperuniform, i.e., $\alpha = 0$.} 
Moreover, we compute the spectral density $\Tilde{\chi}_{_V}(\mathbf{k})$ and use it to characterize the structure of the binary circular-disk packings across length scales and then use it to determine the time-dependent diffusion spreadability of these MRJ-like packings.
The results from this work can be used to inform the experimental design of disordered hyperuniform thin-film materials with tunable degrees of orientational and translational disorder.
\end{abstract}

\maketitle


\section{Introduction}
Disordered hyperuniform media in $d$-dimensional Euclidean space $\mathbb{R}^d$ are exotic states of matter in which volume-fraction fluctuations are anomalously suppressed at infinite wavelengths compared to ordinary liquids \cite{To03, To16, To18_2}.
Such unusual disordered systems have novel physical properties including photonic bandgaps \cite{Fl09,Ma13,Fr17_2,Au20,Kl22}, effective dynamic dielectric and elastic constants \cite{Ki20, To21_3,Vy23}, and fast diffusion spreadabilities \cite{To21_2, Wa22, Sk23}, among many other examples (see, e.g. Ref. \cite{To22}).
Zachary and Torquato \cite{Za09} generalized the hyperuniformity concept to treat heterogeneous two-phase media, of which disordered \textit{packings} are a special case.
Disordered packings---collections of nonoverlapping bodies in $\mathbb{R}^d$---are effective models of physical systems across many disciplines including condensed and soft matter physics \cite{Be65, Ch00, To02, Za07}, materials science \cite{Me94, To02}, and biology \cite{Li01, Pu03, Ge08}, among many others (see Refs. \cite{To10, To18} for additional examples). 
Two-dimensional circular-disk packings in particular have received a great deal of attention \cite{St64,La68,Wo68,Be83,Le86,Ba87,Ko91,Me92,Li93,Gr95,We95,Ja98,Ro99b,Ga06,Mo08,Des10,Hi13,Wu15,Sa17,Tu20,Ko21,Wa21,Am23,Zh23,Ki24_2,Ch24}.
A hyperuniform packing at packing fraction $\phi$ is one in which the local volume fraction variance $\sigma_{_V}^2(R)$ associated with a spherical observation window of radius $R$ decays in the large-$R$ limit faster than $R^{-d}$.
Equivalently, the relevant scattering function---the structure factor $S(\mathbf{k})$ for \textit{only} monodisperse hypersphere packings, and the spectral density $\Tilde{\chi}_{_V}(\mathbf{k})$ otherwise \cite{Za11, Za11_2, Za11_3, Ma22_2}---goes to zero as the wavenumber $k\equiv|\mathbf{k}|$ goes to zero.
Hyperuniform media can be categorized using their hyperuniformity scaling exponents $\alpha > 0$, which describe the scaling of the spectral density in the vicinity of the origin, i.e., $\Tilde{\chi}_{_V}(\mathbf{k})\sim k^{\alpha}$ as $k\rightarrow0$.

Torquato and Stillinger \cite{To03} conjectured that infinitely large packings of identical frictionless hard spheres that are saturated and strictly jammed are hyperuniform.
A packing is strictly jammed if there is no possible collective rearrangement of some finite subset of particles, and no volume nonincreasing deformation can be applied to the packing without violating the impenetrability constraints of the particles \cite{To01,To10}. 
A packing is saturated if there is no available space to add another particle of the same type to the packing without resulting in interparticle overlaps. 
This conjecture has been recently supported by free-volume theory arguments \cite{To21}, which predict that $S(0)=0$ at the jammed state in the absence of defects (i.e., rattlers). While the aforementioned conjecture only applies to monodisperse ordered or disordered sphere packings, subsequent studies have shown this conjecture extends to disordered packings of polydisperse and nonspherical particles \cite{Za11, Za11_2, Za11_3, Ma22_2}.

Maximally random jammed (MRJ) packings of hyperspheres \cite{To00} under the strict jamming constraint have been shown to be hyperuniform \cite{To03, Do05, Sk06, Ho12, At16, Ma22_2, Ma23}.
The MRJ state is the most disordered packing (as measured by some set of scalar order metrics) that is also subject to strict jamming \cite{To00,To10,To18}.
Such states can be viewed as prototypical structural glasses because they are maximally disordered, perfectly rigid (have infinite elastic moduli), and perfectly nonergodic \cite{To10,To18}.
MRJ hypersphere packings are isostatic \cite{Oh03, Do05_2}, which means that the total number of interparticle contacts (constraints) in the packing is equal to the total number of degrees of freedom and that all of the constraints are linearly independent.
This means that for strictly jammed packings of $N$ spheres under periodic boundary conditions, the total number of interparticle contacts is \cite{Do05_2}
\begin{equation}\label{eq:Biiso}
    \bar{Z}=d(N-1) + d(d+1)/2.
\end{equation}

Existing protocols used to produce MRJ-like packings invariably result in packings with a small concentration of \textit{rattlers} $\phi_R$ \cite{Ka02, Sk06, Ji11_3, At13, At14}, which are particles that are not jammed but are locally trapped by their jammed neighbors, which decreases with dimension \cite{Sk06, Ma23}.
The remainder of the jammed spheres are referred to as the \textit{backbone}.
The value of $\phi_R$ for a given $d$ is also known to be affected by the protocol used for jamming \cite{To10_3}.
While one would not generally expect exact hyperuniformity for disordered packings with rattlers, we have shown that when jamming is ensured, the packings are very nearly hyperuniform, and deviations from hyperuniformity correlate with an inability to ensure jamming, suggesting that strict jamming and hyperuniformity are closely linked \cite{At16, Ma23}.

Previous studies of MRJ polydisperse circular-disk and monodisperse sphere packings generated via the Lubachevsky-Stillinger (LS) algorithm \cite{Lu90, Lu91}, conclude that $\alpha = 1$ based on fits of the spectral density \cite{Za11,Za11_2,Foot4} (or structure factor, in the case of Ref. \cite{Do05}), as well as heuristic arguments based on the regularity of the void space between particles \cite{Za11,Za11_2}.
However, recent work on MRJ packings of hyperspheres \cite{Ma23} has demonstrated that ensuring high-quality strict jamming is \textit{required} to extract precise values of $\alpha$ from putatively MRJ packings, which can be achieved using, e.g., the Torquato-Jiao (TJ) algorithm \cite{To10_3}, which is known to reliably produce strictly jammed packings \cite{Ho13, At13, At14, Ma23}.

In this work, we systematically study how the hyperuniformity of polydisperse circular-disk packings is affected by the disk-size distribution.
Due to the infinite size of such a parameter space, we restrict ourselves in this work to binary circular-disk packings (i.e., those with two sizes), with a size ratio $\beta = D_{L}/D_S$, where $D_L$ and $D_S$ are the large and small disk diameters, respectively, and molar ratio of the disk sizes is 1:1.
This restricted parameter space notably contains the well-studied $\beta = 1.4$ equimolar circular-disk system, which is a prototypical structural glass former.

Herein, we use the TJ algorithm to generate a family of strictly jammed disordered packings of equimolar binary circular disks with size ratios $\beta \in [1.01, \textcolor{black}{8.0}]$.
We then characterize ensembles of packings with constant $\beta$ by computing their packing fractions $\phi$, rattler fraction $\phi_R$, fraction of configurations in the ensemble that are isostatic, hyperuniformity scaling exponent $\alpha$, volume fraction variance $\sigma_{_V}^2(R)$, and degree of orientational and translational order.
Subsequently, we compute their spectral densities $\Tilde{\chi}_{_V}(k)$ and use them to characterize the structure of the binary circular-disk packings across length scales, and then use them to determine the time-dependent diffusion spreadability \cite{To21_2} of those packings.

Given the properties of the MRJ state discussed above, we assess which ensembles of packings generated here are MRJ-like according to several criteria, specifically, the rattler fraction $\phi_R$, fraction of configurations for a specific $\beta$ that are isostatic, and the sixfold orientational order metric $\psi_6$.
We find that, for the family of packings with size ratios $1.2\lesssim \beta\lesssim 2.0$, the rattler fraction and orientational order are minimized, and the fraction of packings per constant-$\beta$ ensemble that are isostatic is maximized within the parameter space considered here.
{Per the criteria above, our findings suggest that the aforementioned family of packings is the most MRJ-like of the systems considered herein.
We extract $\alpha$ from from the volume fraction variance $\sigma_{_V}^2(R)$ curves of our packings and find \textcolor{black}{the function $\alpha(\beta)$ is maximized at $\beta$ = 1.4 ($\alpha = 0.450\pm0.002$) within the range $1.2\leq\beta\leq2.0$. Just outside of this range of $\beta$ values, $\alpha(\beta)$ begins to decrease more quickly, and far outside of this range the packings are nonhyperuniform, i.e., $\alpha = 0$.} 
\textcolor{black}{Note that the well-studied case of the $\beta = 1.4$ equimolar disk system happens to produce the maximum $\alpha$ value.}
We use $\sigma_{_V}^2(R)$ to extract $\alpha$ as opposed to $\Tilde{\chi}_{_V}(k)$-based methods because for finite nearly-hyperuniform systems, $\sigma_{_V}^2(R)$ will always have a well-defined hyperuniform scaling regime \cite{To21} from which $\alpha$ can be precisely extracted, which may not be the case with $\Tilde{\chi}_{_V}(k)$-based methods.

The rest of this paper is structured as follows. Section \ref{sec:BiBk} contains pertinent mathematical background and definitions. In Sec. \ref{sec:BiMeth}, we describe the methods used to generate and characterize the MRJ-like binary circular-disk packings. In Sec. \ref{sec:BiRes}, we present the structural characterization of such packings. In Sec. \ref{sec:binspr}, we compute the diffusion spreadability of the jammed binary circular-disk packings. In Sec. \ref{sec:BiCon}, we provide concluding remarks and outlook for future work.

\section{Background}\label{sec:BiBk}
\subsection{Hyperuniformity in Two-Phase Media}\label{sec:BiHU}
A two-phase medium is a partition of space into two disjoint regions called phases \cite{To02}. Let phase one occupy a volume fraction $\phi_1$ and phase two occupy a volume fraction $\phi_2 = 1-\phi_1$.
A packing can be viewed as a two-phase medium, where the space interior to the particles is one phase, and the space exterior is another.
In this subsection, phase two is the volume taken up by the disk packing, so $\phi_2$ is equivalent to the packing fraction $\phi$.
The two-phase medium can be fully statistically characterized by the $n$-point correlation functions \cite{To02}
\begin{equation}\label{BiSfunc}
    S_n^{(i)}(\textbf{x}_1,\ldots,\textbf{x}_n) \equiv \braket{\mathcal{I}^{(i)}(\textbf{x}_1) \dots \mathcal{I}^{(i)}(\textbf{x}_n)}, 
\end{equation}
\noindent where $\mathcal{I}^{(i)}(\textbf{x})$ is the indicator function of phase $i = 1,2$, and the angular brackets indicate an ensemble average. The function $S_n^{(i)}(\textbf{x}_1,\ldots,\textbf{x}_n)$ gives the probability of finding the vectors $\textbf{x}_1,\ldots,\textbf{x}_n$ all in phase $i$. The autocovariance function $\chi_{_V} (\textbf{r})$ is related to the two-point correlation function $S_2^{(i)}(\mathbf{r})$ by
\begin{equation} \label{Biauto}
    \chi_{_V}(\textbf{r})\equiv S_2^{(1)}(\textbf{r}) - \phi_1^2 = S_2^{(2)}(\textbf{r}) - \phi_2^2, 
\end{equation}
assuming statistical homogeneity. 
The nonnegative spectral density $\Tilde{\chi}_{_V}(\mathbf{k})$, which can be obtained via scattering experiments \cite{De57}, is the Fourier transform of $\chi_{_V}(\textbf{r})$.
For statistically homogeneous and isotropic media, the slope of $\chi_{_V}(\mathbf{r})$ at the origin is directly related to the specific surface $s$, which is the interface area per unit volume. Specifically, in any space dimension $d$, it is known that the small-$\mathbf{|r|}$ asymptotic form is given by \cite{To02}
\begin{equation}
    \chi_{_V}(\mathbf{r})=\phi_1\phi_2 - \eta(d)s|\mathbf{r}|+\mathcal{O}(|\mathbf{r}|^2),
\end{equation}
where
\begin{equation}
    \eta(d)=\frac{\Gamma(d/2)}{2\sqrt{\pi}\Gamma((d+1)/2)}.
\end{equation}
The specific surface $s$ is a useful characteristic microscopic length scale associated with two-phase media.

The local volume-fraction variance $\sigma_{V}^2(R)$ associated with a sampling window of radius $R$ can be written in terms of $\chi_{_V}(\textbf{r})$  as \cite{Lu90_2,Za09}
\begin{equation}\label{Bivolumefractionreal}
    \sigma_{V}^2(R) = \frac{1}{v_1(R)} \int_{\mathbb{R}^d}  \chi_{_V}(\textbf{r})\alpha_2(r;R)d\textbf{r},
\end{equation}
and in terms of $\Tilde{\chi}_{_V}(\mathbf{k})$ as \cite{Za09, To18_2}
\begin{equation}\label{Bivolume_fromspec}
    \sigma_{V}^2(R) = \frac{1}{v_1(R)(2\pi)^d} \int_{\mathbb{R}^d}  
\Tilde{\chi}_{_{V}}(\textbf{k}) \Tilde{\alpha}_2(k;R)d\textbf{k},
\end{equation}
where 
\begin{equation}
    v_1(R) = \frac{\pi^{d/2}R^d}{\Gamma(d/2 + 1)},
\end{equation}
is the volume of a $d$-dimensional sphere, $\Gamma(x)$ is the gamma function, $\alpha_2(r;R)$ is the intersection volume of two spherical windows of radius $R$ separated by a distance $r$ scaled by the volume of one window, and $\Tilde{\alpha}_2(k;R)$ is its Fourier transform. 

A hyperuniform two-phase medium has $\sigma_{_V}^2(R)$ that decays more quickly than the window volume $R^{-d}$ in the large-R regime \cite{Za09, To18_2}, i.e., 
\begin{equation}
    \lim_{R\rightarrow\infty}R^d\sigma_{_V}^2(R) = 0.
\end{equation}
Equivalently, a hyperuniform medium has a spectral density that goes to zero as the wavenumber goes to zero \cite{Za09, To18_2}, i.e.,
\begin{equation}
    \lim_{|\mathbf{k}|\rightarrow 0}\Tilde{\chi}_{_V}(\mathbf{k}) = 0.
\end{equation}
Consider a hyperuniform system whose $\Tilde{\chi}_{_V}(\mathbf{k})$ has the following power-law behavior as $|\mathbf{k}|$ tends to 0:
\begin{equation}
    \Tilde{\chi}_{_V}(\mathbf{k})\sim |\mathbf{k}|^{\alpha}\quad(|\mathbf{k}|\rightarrow0),
\end{equation}
where $\alpha>0$ is the hyperuniformity scaling exponent.
It has been shown \cite{Za09, To18_2} that there are three distinct scaling regimes (classes) that describe the associated large-$R$ behaviors of the volume-fraction variance
\begin{numcases}{\sigma_{_V}^2(R) \sim }\label{BiV_classesI}
  R^{-(d+1)}, \quad \alpha >1  & \(  (\text{Class I})\) \nonumber \\
  R^{-(d+1)}\ln(R),\quad \alpha = 1 & \( (\text{Class II})\)   \\
  R^{-(d+\alpha)}, \quad (0 < \alpha < 1) & \( (\text{Class III}).\)   \nonumber
\end{numcases}
Classes I and III are the strongest and weakest forms of hyperuniformity, respectively.
Notably, for class II and III systems, one can extract a precise value of $\alpha$ from the large-$R$ scaling of $\sigma_{_V}^2(R)$, but this is not possible for class I systems that all have the same large-$R$ scaling regardless of $\alpha$ \cite{To18_2}.

For a large class of statistically homogeneous two-phase media in $\mathbb{R}^d$, the large-$R$ asymptotic expansion of $\sigma_{_V}^2(R)$ is given by \cite{Za09}
\begin{equation}\label{eq:BiAsy}
    \sigma_{_V}^2(R) = \bar{A}_{_V} \left(\frac{D}{R}\right)^d + \bar{B}_{_V} \left(\frac{D}{R}\right)^{d+1} + o\left(\frac{D}{R}\right)^{d+1},
\end{equation}
where $o(x)$ signifies terms of order less than $x$, $D$ is a characteristic microscopic length scale of the medium, and $\bar{A}_{_V}$ and $\bar{B}_{_V}$ are, respectively, asymptotic coefficients that multiply terms proportional to the window volume and window surface area.
When $\bar{A}_{_V}=0$, $\bar{B}_{_V}$ must be positive, implying that the medium is hyperuniform of class I [cf. Eq. (\ref{BiV_classesI})].
Torquato \textit{et al}. \cite{To22_2} showed that $\bar{B}_{_V}$ can be used to quantify the large-scale order of hyperuniform media, where larger values of $\bar{B}_{_V}$ indicate the medium is more disordered at large length scales.

\subsection{Orientational Order Metric}
To assess bond orientational order in disk packings, we use $\psi_n$, which measures how closely the local environment around each disk in the packing resembles perfect $n$-fold symmetry \cite{Ha78, Ne79, Ka00}.
This order metric is defined as
\begin{equation}\label{eq:Bipsi}
    \psi_n = \frac{1}{N}\sum_{k=1}^N\frac{1}{N_{S,k}}\left|\sum_j^{N_{S,k}}e^{ni\theta_{k,j}} \right|,
\end{equation}
where $N$ is the number of disks in the packing, $N_{S,k}$ is the number of ``bonds'' that disk $k$ has to its neighboring disks (equal to the number of edges on its corresponding Voronoi polygon), and $\theta_{k,j}$ is the angle between some arbitrary reference direction and the vector connecting disks $k$ and $j$.
Such a metric takes values between zero and one where one indicates that each particle in the system has perfect $n$-fold symmetry.

\subsection{Spreadability}
The diffusion spreadability, introduced by Torquato in Ref. \cite{To21_2}, is a dynamical probe that directly links the time-dependent diffusive transport with the microstructure of heterogeneous media across length scales.
Consider the time-dependent problem describing the mass transfer of a solute between two phases and assume that the solute is initially only present in one phase, specifically the particle phase, and both phases have the same $\mathcal{D}$.
The fraction of total solute present in the void space as a function of time $\mathcal{S}(t)$, is termed the \textit{spreadability} because it is a measure of the spreadability of diffusion information as a function of time.
Qualitatively, given two different two-phase systems at some time $t$, the one with a larger value of $\mathcal{S}(t)$ spreads diffusion information more rapidly.
Recently, Torquato showed that the \textit{excess spreadability} $\mathcal{S}(\infty)-\mathcal{S}(t)$ can be expressed in Fourier space in any dimension $d$ as \cite{To21_2}
\begin{equation}\label{eq:BiFT_spread}
    \mathcal{S}(\infty)-\mathcal{S}(t) = \frac{dv_1(1)}{(2\pi)^d \phi}\int^{\infty}_0 k^{d-1}\spD{k}\textrm{exp}[-k^2\mathcal{D}t]dk.
\end{equation}
The small-, intermediate-, and long-time behaviors of the excess spreadability enable one to probe the small-, intermediate, and large-scale characteristics of the microstructures, respectively \cite{To21_2}.
Recent work has utilized the spreadability to probe the microstructures of standard disordered media \cite{To21_2, Wa22, Sk23}, ordered and disordered hyperuniform media \cite{To21_2, Wa22, Ma22_2, Sk23} (including self-similar quasicrystalline media \cite{Hi24}), and experimental samples of geological \cite{Wa22} and biological media \cite{Li24} across length scales.

\section{Packing Generation and Characterization Methods}\label{sec:BiMeth}
\subsection{Torquato-Jiao Algorithm}
To generate the packings examined in this work, we employ the Torquato-Jiao (TJ) algorithm \cite{To10_3}.
In this algorithm, the problem of generating dense packings of $N$ nonoverlapping particles within a fundamental cell in $\mathbb{R}^d$, subject to periodic boundary conditions, is formulated as an optimization problem called the adaptive shrinking cell scheme \cite{To09}, where the objective function is taken to be the negative of the packing fraction and minimized using linear-programming (LP) techniques, since the design variables become exactly linear as jammed states are approached.
Starting from an initially unsaturated packing configuration of $d$-dimensional hyperspheres of a fixed size in the fundamental cell, the positions of the particles are the design variables for the optimization.
We allow the boundary of the fundamental cell to deform macroscopically as well as compress or expand (while keeping the particles fixed in size) such that the volume of the simulation cell decreases on average, also making the box shape and size design variables in the optimization.
The TJ algorithm is able to generate strictly jammed packings to within a small numerical tolerance \cite{To10_3}.

Here, we start by randomly placing nonoverlapping binary disks with a size ratio $\beta$ and a 1:1 molar ratio in a square simulation box with an initial packing fraction of 0.4.
We use the same simulation parameters as those in Ref. \cite{Ma23}, which were shown to generate high-quality strictly jammed hypersphere packings for $d = 3,4,5$.
In particular, we use an influence sphere size $\gamma = D/40$, a maximum translation magnitude of $\Delta x\leq D/200$, and a maximum box strain magnitude of $\epsilon\leq D/200$.
Simulations are terminated when the packing fraction $\phi$ decreases less than $10^{-15}$ over the course of two LP steps.

We generated 1000-packing ensembles of disks with size ratios of $\beta =$ 1.01, 1.05, 1.1, 1.15, 1.2, 1.25, 1.3, 1.35, 1.4, \textcolor{black}{1.45, 1.5, 1.55}, 1.6, \textcolor{black}{1.65, 1.7, 1.75,} 1.8, \textcolor{black}{1.9,} 2.0, 2.5, 3.0 \textcolor{black}{4.0, 5.0, 6.0, 7.0, and 8.0}.
We find that for system sizes greater than 2500 particles, we are unable to reach high-quality jammed states as outputs of the TJ algorithm, and thus, use $N = 2500$ for all packings examined in this work.
\textcolor{black}{Low-quality jammed states with large numbers of particles have been shown to have nonhyperuniform scaling \cite{Ma23}, such nonhyperuniform scaling would thus occur in our packings if $N > 2500$ disks were used.}
\textcolor{black}{The number of disks used here, when accounting for space dimension, is larger than the number of spheres used in the 3D, 4D, and 5D MRJ hypersphere packings examined in previous work \cite{Ma23} and thus can reliably be used to study the hyperuniformity of such systems.}
Illustrative examples of the packings that this method produces are shown in Fig. \ref{fig:bin_configs}.

\begin{figure}[!]
    \centering
    \subfigure[]{\includegraphics[height=0.34\textwidth]{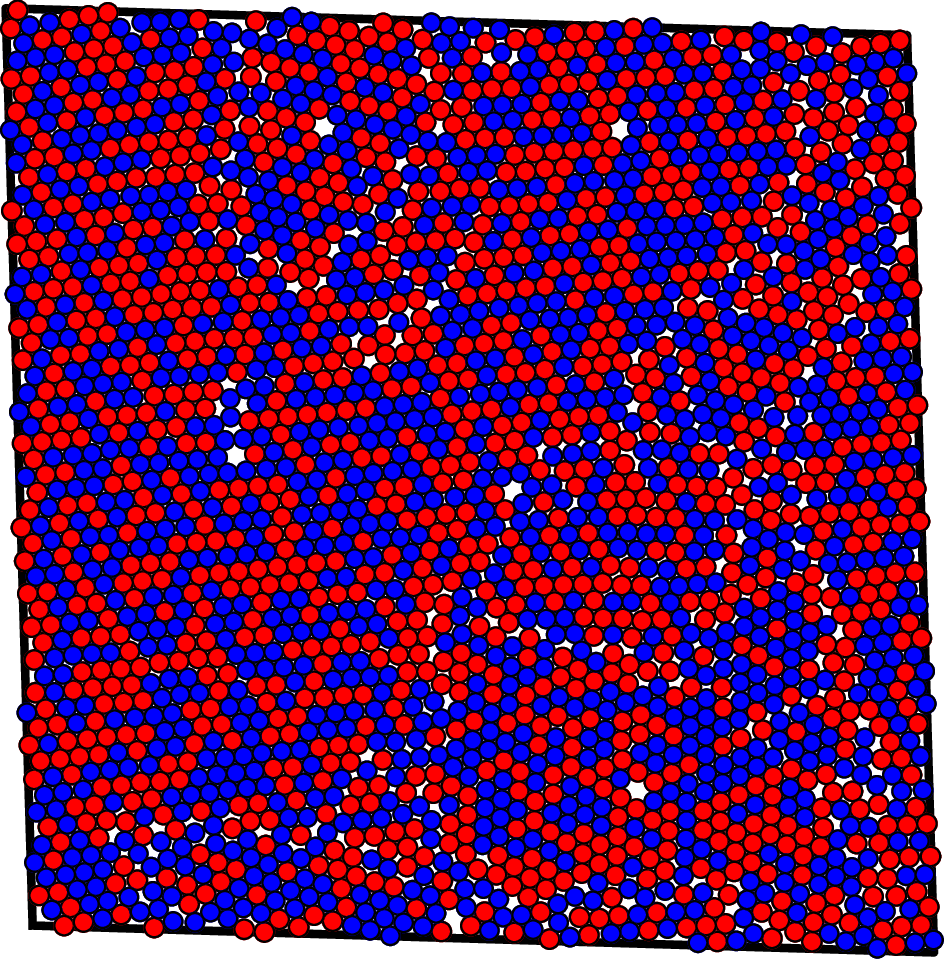} }
    \subfigure[]{\includegraphics[height=0.34\textwidth]{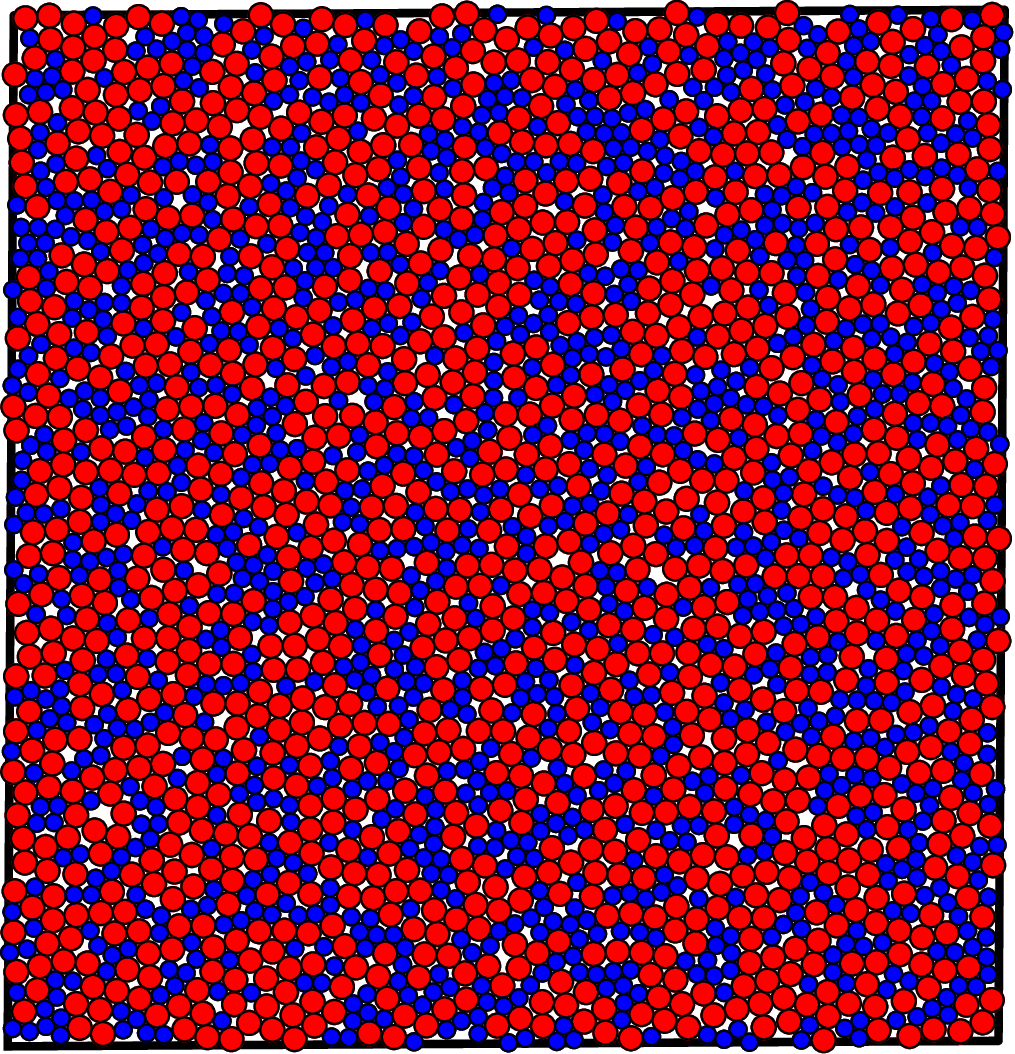} }
    \subfigure[]{\includegraphics[height=0.34\textwidth]{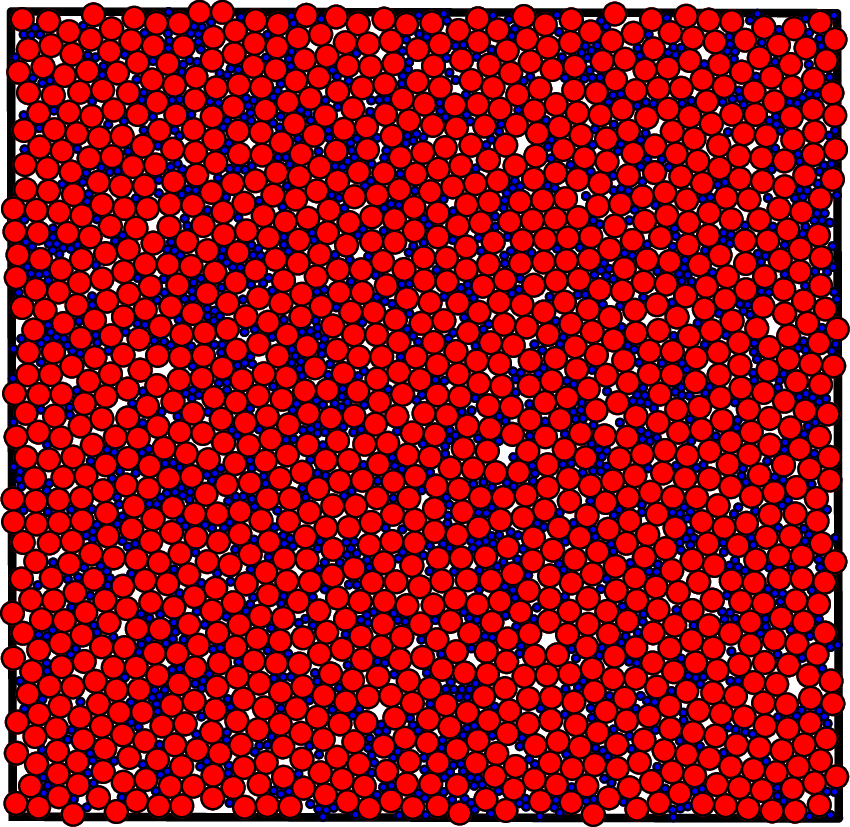} }
    \caption{Representative configurations of strictly jammed binary disk packings with $N = 2500$ with size ratios (a) $\beta= 1.05$, (b) $\beta=1.4$, and (c) $\beta=3.0$. Note that the TJ algorithm allows the simulation box to deform over the course of the simulation, which is why the top panel has a clearly skewed box shape, unlike the other two examples.}
    \label{fig:bin_configs}
\end{figure}

\subsection{Isostaticity}
In numerically generated packings, no two spheres are ever exactly contacting each other, so we treat two disks as contacting if they are within a small distance of each other called the \textit{contact tolerance}.
To determine $\phi_R$ and the fraction of configurations in each ensemble with different size ratios $\beta$ that are isostatic\textcolor{black}{, we follow the procedure used in Ref. \cite{Ma23}.}
\textcolor{black}{In brief, we choose a contact tolerance, then iteratively determine the number of contacts for each disk, then remove any disk that is a rattler (i.e., that does not have at least three non-cohemispheric contacts) until no rattlers are removed.}
\textcolor{black}{At this point, we have determined the set of disks that make up the jammed backbone of the packing, which allows us to determine the average contact number and rattler fraction for a given contact tolerance.} 
We repeat this process for several contact tolerances between $10^{-14}\langle D\rangle $ and $10^{-2}\langle D\rangle $, where $\langle D\rangle$ is the average disk diameter.
\textcolor{black}{We repeat this process for a broad range of contact tolerances because previous work \cite{Do05_2,Ma23} has demonstrated that in high-quality jammed packings the average contact number $\bar{Z}$ will be constant over several orders of magnitude of the contact tolerance.}
For each packing, we find this plateau and then determine corresponding $\bar{Z}$ and $\phi_R$ values.
If $\bar{Z}$ matches the value from Eq. (\ref{eq:Biiso}), then it is counted as isostatic.
Rattler fraction $\phi_R$ values are ensemble averaged over the 1000 configurations for each size ratio.
\textcolor{black}{We note here that rattlers are removed \textit{only} when computing $\bar{Z}$ and are left in for all other calculations described in this section.}
\textcolor{black}{Such a method is consistent with previous studies of MRJ sphere packings where removing the ratters has been shown to destroy hyperuniformity \cite{Do05}.}

\subsection{Orientational Order}
To compute the orientational order metrics $\psi_n$, we use the implementation from the {\tt freud} library \cite{Fr20}.
In short, the Voronoi diagram for each packing is computed and then used with Eq. (\ref{eq:Bipsi}) to determine $\psi_n$ for each packings.
In particular, we compute $\psi_n$ for $n=5,6,7$ and 8 and ensemble average these values for each of the 1000-packing ensembles.

\subsection{Local Volume-Fraction Variances}
We compute $\sigma_{_V}^2(R)$ by directly sampling the volume fraction of the disks within randomly placed windows of radius $R$ \cite{At16,Za11,Za11_2,Za09,To22_2,Ki21,Sk24}.
We choose the number of sampling windows per configuration using the method described in Ref. \cite{Kl21_2}. 
We then determine the power-law scaling of $\sigma_{_V}(R)$ in the large-$R$ limit via a linear regression on the large-$R$ region of $\sigma_{_V}^2(R)$ on a logarithmic scale and from this power-law scaling we can determine $\alpha$ by comparing this power-law scaling to Eq. (\ref{BiV_classesI}).
The coefficient on this power-law scaling can then be used to determine the surface-area scaling coefficient $\bar{B}_{_V}$ [cf. Eq. (\ref{eq:BiAsy})], which we use to quantify the large-scale translational order in these packings.
We choose to extract $\alpha$ using $\sigma_{_V}^2(R)$ as opposed to directly fitting the small-$k$ region of $\Tilde{\chi}_{_V}(k)$ or the recently developed spreadability methods \cite{To21_2,Wa22, Ma22_2,Sk23,Ma23,Hi24} because, based on arguments presented in Ref. \cite{To21}, variance-based structural characterizations of finite or nearly hyperuniform systems will always exhibit well-defined hyperuniform scaling regimes from which $\alpha$ can be precisely extracted, which is not guaranteed when using the direct-fitting or spreadability methods mentioned above.

\subsection{Spectral Densities}
To compute $\Tilde{\chi}_{_V}(\mathbf{k})$ for these packings, we use the direct Fourier transform of the disks in the packing given by \cite{Za11_2}
\begin{equation}
    \Tilde{\chi}_{_V} = \frac{|\sum_{j=1}^N\text{exp}(-i\mathbf{k}\cdot{\mathbf{r_j}})\Tilde{m}(\mathbf{k};R_j)|^2}{V}\quad(\mathbf{k}\neq\mathbf{0}),
\end{equation}
where $\mathbf{r}_j$ is the position of the centroid of disk $j$, $V$ is the volume of the simulation box, and $\Tilde{m}(\mathbf{k};R_j)$ is the Fourier transform of the indicator function for disk $j$ (see Sec. \ref{sec:BiHU}).
The radially-averaged $\Tilde{\chi}_{_V}(k)$ are ensemble averaged for each of the 1000-packing ensembles.
We then substitute these $\Tilde{\chi}_{_V}(k)$ into Eq. (\ref{eq:BiFT_spread}) to compute the excess spreadabilities $\mathcal{S}(\infty) - \mathcal{S}(t)$ for each ensemble.

\section{Results and Discussion}\label{sec:BiRes}
Here, we present the structural characterization of the packings generated with the TJ algorithm (per the methods described in Sec. \ref{sec:BiMeth}) and subsequent analysis of these characterizations.
In particular, we first examine how the packing fraction $\phi$, rattler fraction $\phi_R$, fraction of configurations in an ensemble with fixed $\beta$ that are isostatic, and orientational order vary as a function of the size ratio $\beta$.
Then, we compute $\sigma_{_V}^2(R)$ to determine how the hyperuniformity scaling exponents $\alpha(\beta)$ and window-surface-area coefficients $\bar{B}_{_V}$ change with $\beta$.
Last, we present the spectral densities $\Tilde{\chi}_{_V}(k)$.

\begin{figure}[!t]
    \centering
    \includegraphics[width=0.45\textwidth]{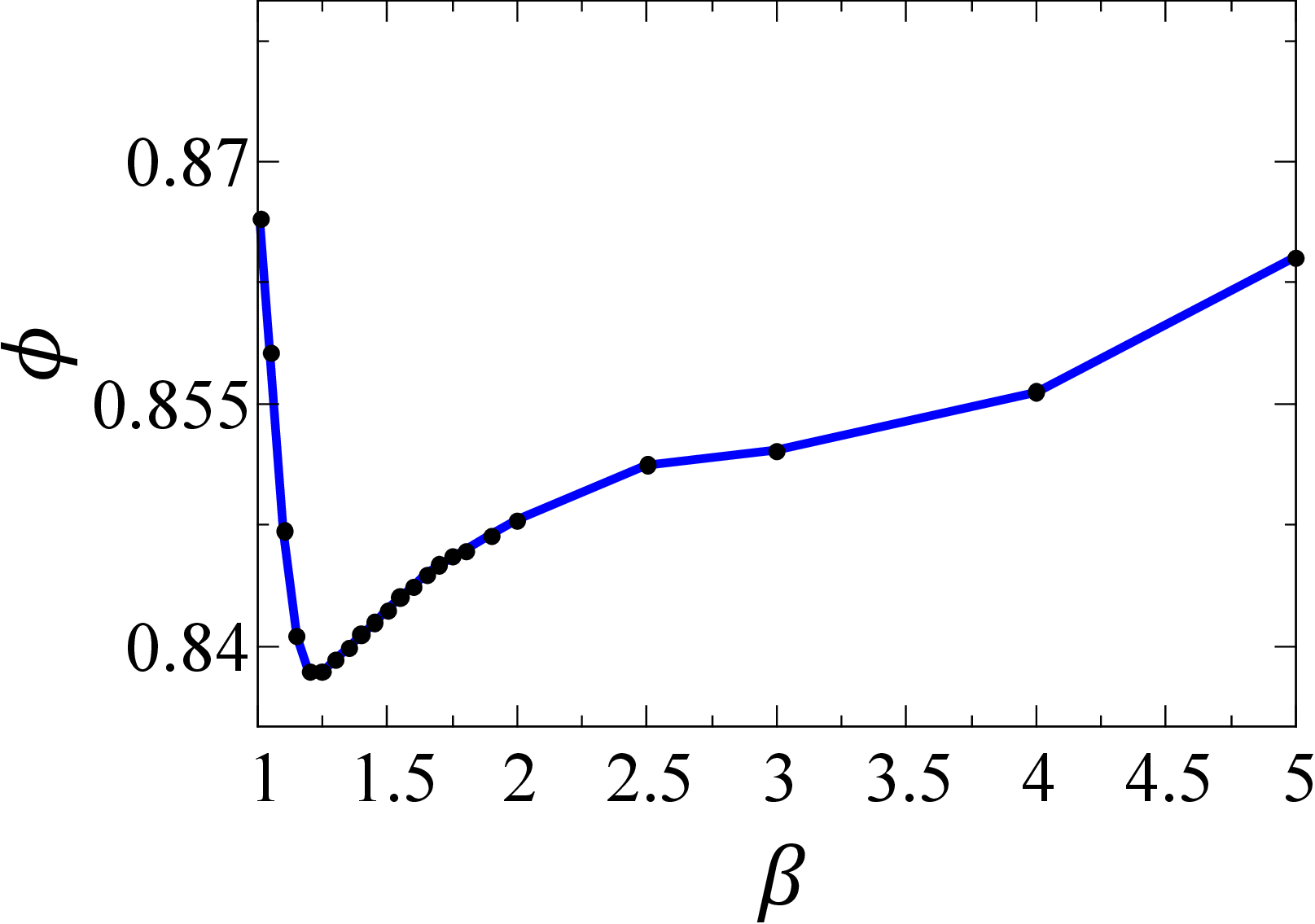}
    \caption{Ensemble averaged packing fraction $\phi$ for selected values of the size ratio $\beta\in[1.01,5.0]$ for the strictly jammed binary disk packings. Black dots correspond to the data points and the blue connecting lines are drawn for visual purposes.
    }
    \label{fig:bin_phi}
\end{figure}

\begin{figure}[!t]
    \centering
    \includegraphics[width=0.45\textwidth]{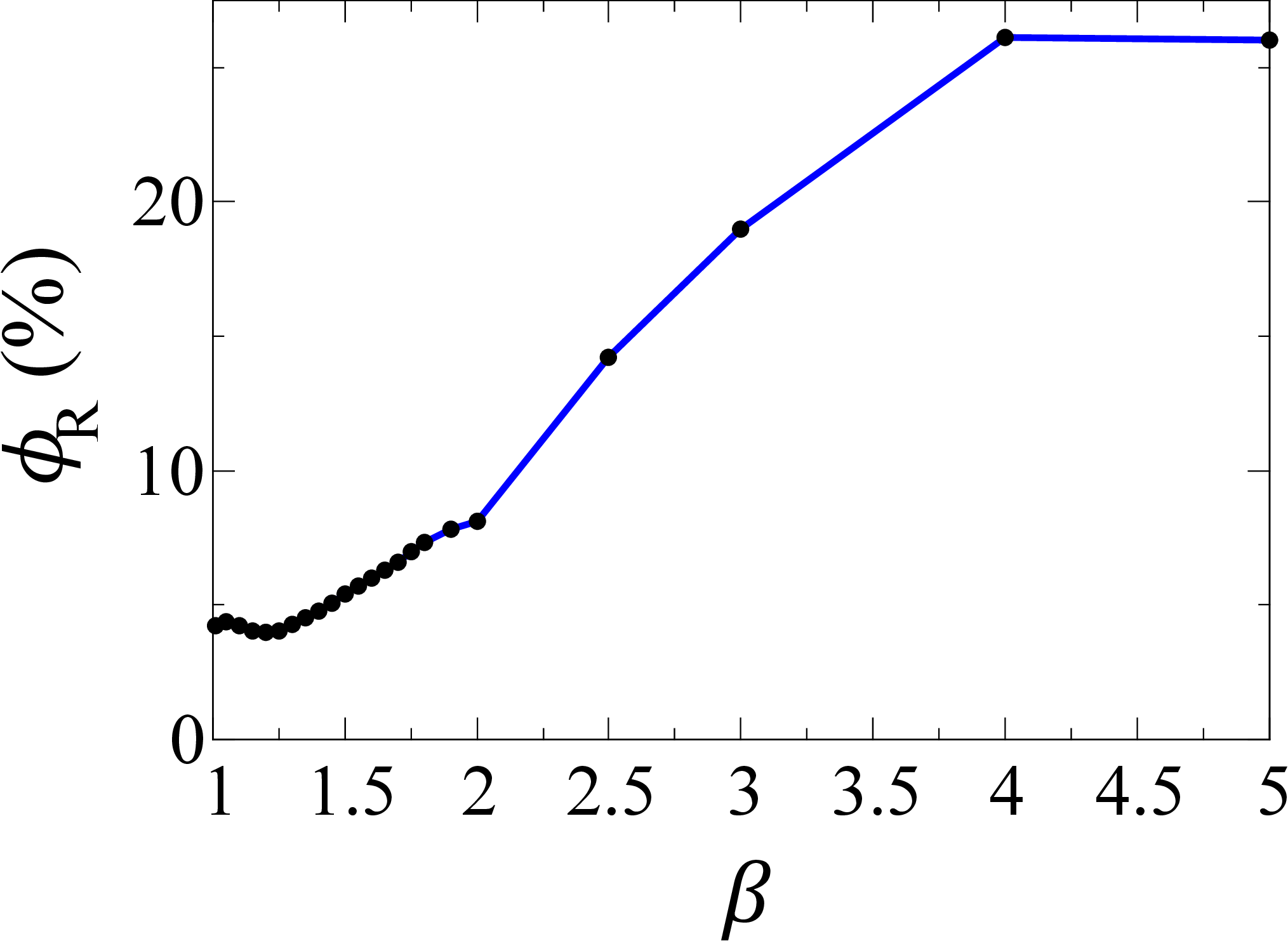}
    \caption{Ensemble averaged percentage rattler fraction $\phi_R$ for selected values of the size ratio $\beta\in[1.01,5.0]$ for the strictly jammed binary disk packings. Black dots correspond to the data points and the blue connecting lines are drawn for visual purposes.
    }
    \label{fig:bin_phiR}
\end{figure}

\subsection{Isostaticity}
Recall that MRJ states are isostatic \cite{Oh03, Do05_2}, which is an important criterion to be the most disordered jammed packing, as increasing or decreasing $\bar{Z}$ while maintaining strict jamming requires an increase in the order of the packing \cite{To10_3, Ji11_3}.
Figures \ref{fig:bin_phi} and \ref{fig:bin_phiR} show the packing fraction and rattler fractions for each ensemble of disk packings as a function of $\beta$.
The raw values for $\phi$ and $\phi_R$ are given in Table \ref{tab:bin}.
Figures \ref{fig:bin_phi} shows a clear trough in the $\phi$ as a function of $\beta$, with a minimum of $\phi = 0.8384$ at $\beta = 1.2$ and 1.25.
The packing fraction increases as $\beta$ decreases due to the packings becoming increasingly polycrystalline (supported also by an increase in $\psi_6$ with decreasing $\beta$; see below) and also increases as $\beta$ increases because the smaller disks are able to more easily fit within the gaps between the larger disks, which increases $\phi$.
The rattler fraction also exhibits a minimum of $\phi_R = 3.99\%$ at $\beta = 1.2$.
For $\beta < 1.2$ the rattler fraction holds roughly constant at $\phi_R\approx 4\%$, but increases as $\beta$ increases because the small disks that lie in the gaps between the large disks tend to be rattlers.
We note that the rate at which $\phi_R$ increases with $\beta$ increases at $\beta \gtrsim 2.0$.
\textcolor{black}{Similarly, we find that the fraction of configurations in an ensemble with fixed $\beta$ that are isostatic is peaked at 38.4\% when $\beta = 1.3$, is greater than $\sim10\%$ for $\beta \lesssim 2.0$, and then is $<$1\% for $\beta > 2.0$ (cf. Table \ref{tab:bin}).}

\begin{figure}[!t]
    \centering
    \includegraphics[width=0.45\textwidth]{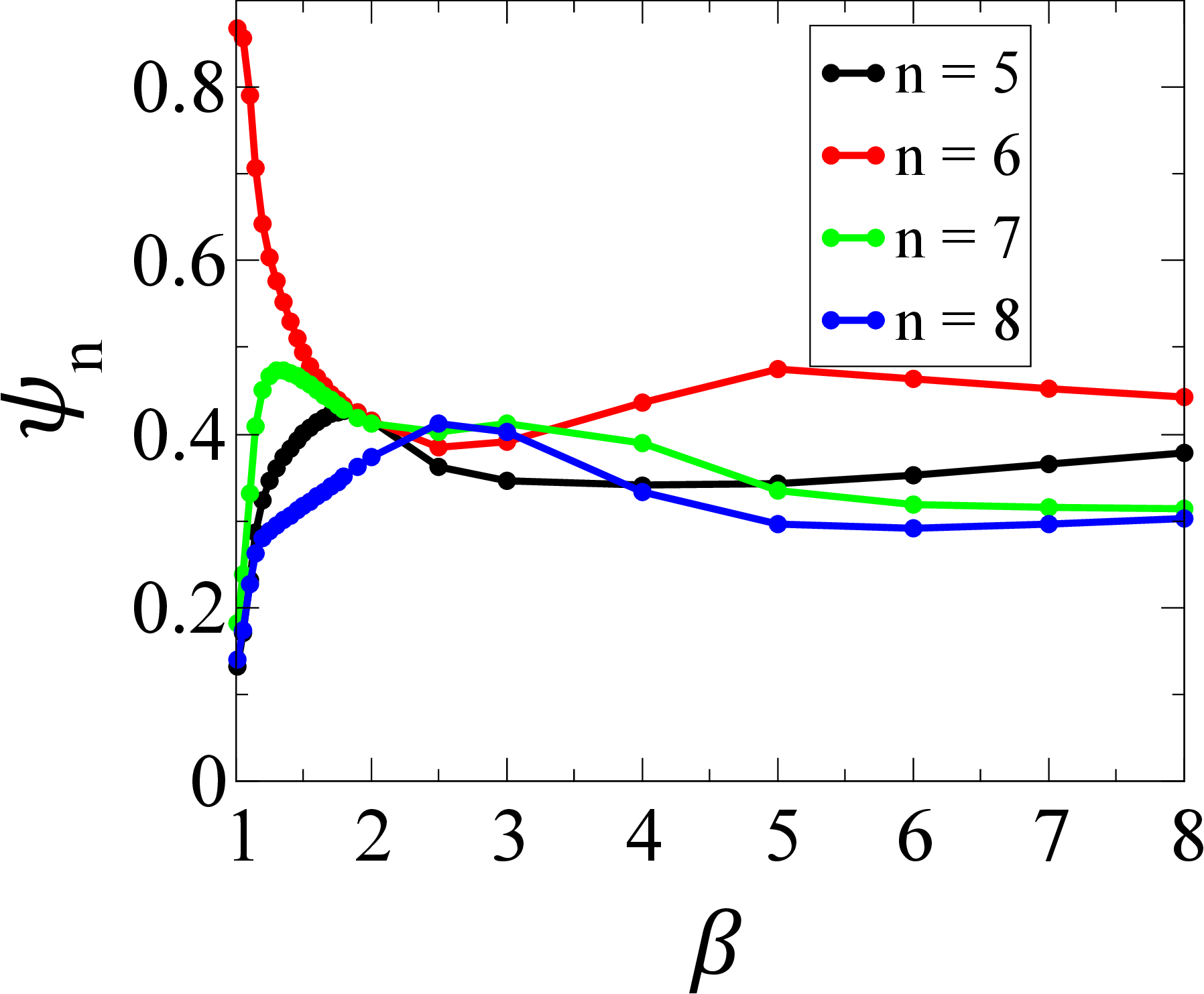}
    \caption{The ensemble-averaged $\psi_n$ orientational order metric for $n = 5,6,7,$ and 8 for selected values the size ratio $\beta\in[1.01,5]$ for the strictly jammed binary disk packings.
    Solid dots correspond to data points and the connecting lines are drawn for visual purposes.
    }
    \label{fig:bin_Qn}
\end{figure}

\subsection{Orientational Order}
Figure \ref{fig:bin_Qn} shows the orientational order metrics $\psi_n$ for $n = 5,6,7,8$ as a function of $\beta$.
At small $\beta$, the packings have a large number of crystallites with sixfold rotational symmetry, which is supported by the large value of $\psi_6$, and the relatively small values of $\psi_{5,6,8}$.
As $\beta$ increases from 1, the number of local neighborhoods with fivefold, sevenfold, and eightfold orientational order rises sharply.
At $\beta\approx1.2$, $\psi_{5,7,8}$ begin to increase more slowly and at $\beta\approx1.35$, the number of local neighborhoods with severfold orientational symmetry begins to decrease, while those with fivefold and eightfold symmetry continue to increase.
As $\beta$ increases, we find that $\psi_{5,6,7,8}$ begin to \textcolor{black}{plateau between 0.3 and 0.4}, which indicates that the sixfold orientational order associated with the perfect hexagonal crystal becomes less common and local neighborhoods with other orientational symmetries (i.e., those not associated with the perfect hexagonal crystal) become more common.

\subsection{Family of MRJ-like Binary Packings}
Recall that the MRJ state is defined as the most random state (as measured by a set of correlated scalar order metrics) subject to strict jamming \cite{To00}.
Moreover, recall that MRJ hypersphere packings are isostatic \cite{Oh03, Do05_2} and---in the ideal case---rattler-free \cite{At16, To21}.
Thus, here we will use the sixfold orientational order metric $\psi_6$, $\phi_R$, and the fraction of packings in a constant-$\beta$ ensemble that is isostatic as the criteria by which we determine which packings are MRJ-like.
The strictly jammed MRJ state for \textit{monodisperse} disks was found to have $\psi_6 = 0.663\pm0.02$ \cite{At14}.
Using this as a baseline for our strictly jammed binary disk packings, we note that all binary disk packing ensembles with $\beta\geq1.2$ have $\psi_6\leq0.663$ (cf. Fig. \ref{fig:bin_Qn}), i.e., have greater orientational disorder than the monodisperse MRJ state.
In addition, we find that there is a rapid increase in the rattler concentration and rapid drop-off in the fraction of isostatic configurations for ensembles with $\beta>2.0$.
Taking the intersection of the set of ensembles with orientational disorder greater than that of existing MRJ 2D disk packings ($\beta\gtrsim1.2$) and the set of ensembles with a high fraction of isostaticity and low fraction of rattlers relative to the rest of the ensembles ($\beta\lesssim2.0$), we conclude that the family of jammed binary equimolar disk packings with $1.2\leq\beta\leq2.0$ have structural characteristics that make them the most MRJ-like of the packings considered here.
Henceforth, we refer to this family of MRJ-like packings as being in the ``MRJ-like size ratio regime.''

\begin{figure}
    \centering
    \includegraphics[width=0.4\textwidth]{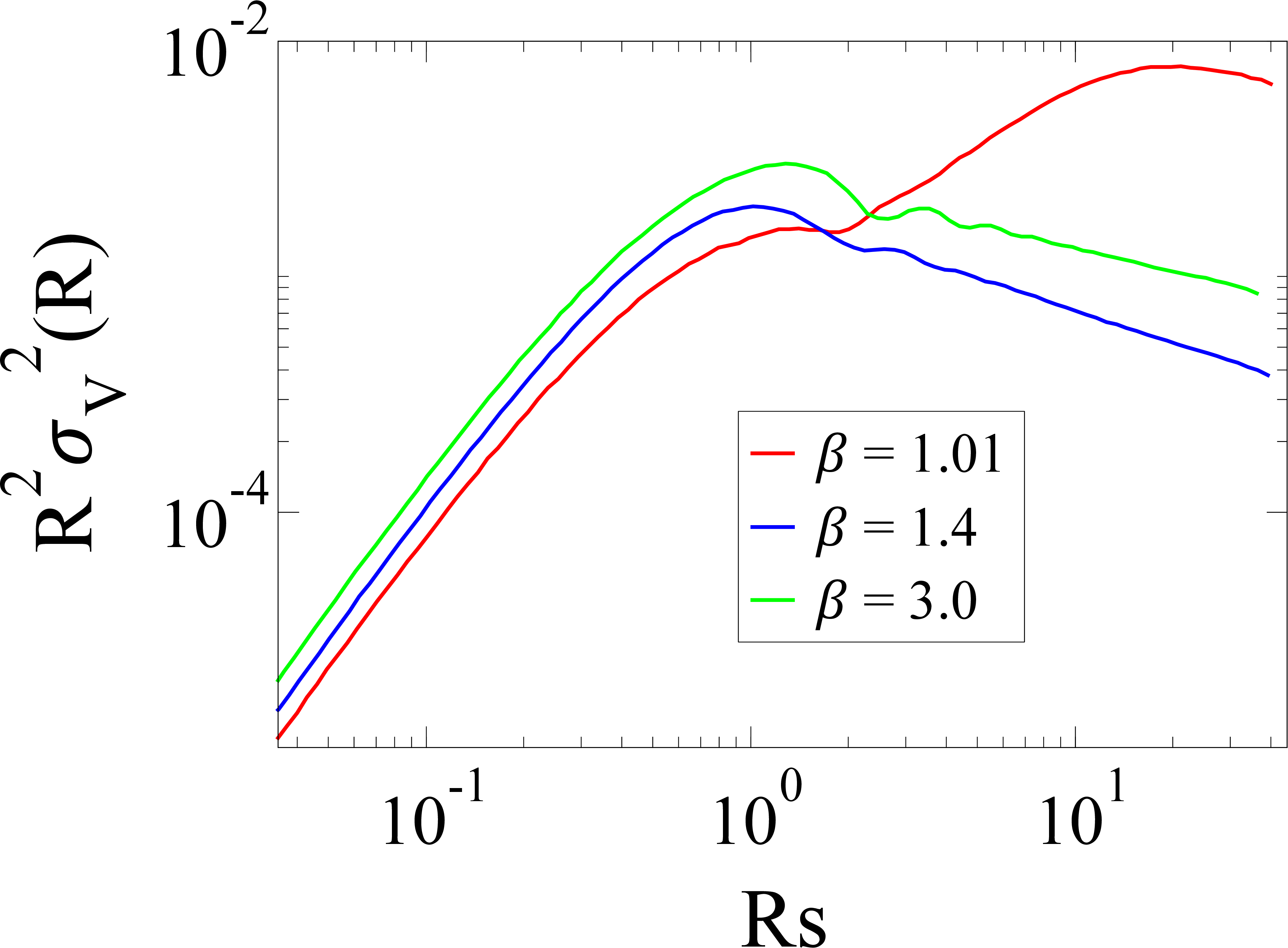}
    \caption{Representative curves of the volume fraction variance $\sigma_{_V}^2(R)$ scaled by the sampling window volume $R^2$ as a function of window radius scaled by the specific surface $Rs$. 
    }
    \label{fig:binvv}
\end{figure}

\subsection{Local Volume-Fraction Variances}
Figure \ref{fig:binvv} shows representative $\sigma_{_V}^2(R)$ curves scaled by the window volume $R^2$ for packings with the smallest $\beta$ considered in this work, a large $\beta$ value outside of the MRJ-like regime, and $\beta$ in the MRJ-like regime.
Two-dimensional media whose $R^2\sigma_{_V}^2(R)$ decay at large-$R$ are hyperuniform (see Sec. \ref{sec:BiHU}).
The $\beta = 1.4$ and 3.0 curves in Fig. \ref{fig:binvv} have stable decay rates at large-$R$, indicating that they are hyperuniform.
We then fit the large-$R$ regime of $\sigma_{_V}^2(R)$ to extract precise values of $\alpha$ and calculate the window surface-area coefficient $\bar{B}_{_V}$, the values of which are presented in Table \ref{tab:bin}.
We scale $\bar{B}_{_V}$ by the specific surface $s$, which for these binary disk packings is given by \cite{To02}
\begin{equation}\label{eq:BinSS}
    s(\beta) = 2\phi\frac{1+\beta}{1+\beta^2},
\end{equation}
where the radius of the small disks is set to 1.
Here, and in subsequent sections, we use the inverse of the specific surface $s^{-1}$ as a characteristic microscopic length scale to make the window radius $R$, the wavenumber $k$, and window-surface area coefficient $\bar{B}_{_V}$ dimensionless, since the specific surface is a convenient and easily measured quantity for general two-phase microstructures, as advocated in Refs. \cite{To22_2, Ki21}.
The uncertainty on the measurements of $\alpha$ and $\bar{B}_{_V}$ come from the standard error of the linear regression used to compute $\alpha$ and $\bar{B}_{_V}$.

We find that for $\beta \leq 1.1$ \textcolor{black}{and $\beta>4$} the large-$R$ scaling regime does not have a stable decay and thus can be considered nonhyperuniform. Spurious increases in decay rate at large $R$, such as the one observed in the $\beta=1.01$ curve in Fig. \ref{fig:binvv} have been shown to result from finite size effects \cite{At16}}.  
\textcolor{black}{Fig. \ref{fig:avb} shows the values of the hyperuniformity scaling exponent $\alpha(\beta)$ as a function of the size ratio $\beta$ for the hyperuniform packing ensembles.}
At $\beta = 1.4$, we find \textcolor{black}{a maximum} $\alpha = 0.450\pm0.002$, which is consistent with recent analyses of hyperuniform RCP packings of binary disks with $\beta = \sqrt{2}=1.41...$ \cite{Wi21, Wi23_2}. 
\textcolor{black}{Note that the well-studied case of the $\beta = 1.4$ equimolar disk system happens to produce the maximum $\alpha$ value.}
\textcolor{black}{Within the MRJ-like regime, $\alpha(\beta)$ decreases linearly as $\beta$ deviates from $\beta = 1.4$. In particular, $\alpha(\beta) = 0.2704 + 0.126\beta$ for $1.2\leq\beta\leq1.4$ and $\alpha(\beta) = 0.5436-0.0683\beta$ for $1.4\leq\beta\leq2.0$.}
\textcolor{black}{Outside of the MRJ-like range, $\alpha$ begins to decrease more quickly and ultimately vanishes for large $\beta$ or $\beta$ close to 1, which can be attributed to an increase in $\phi_R$ for large $\beta$, and an increase in crystallite concentration for small $\beta$.}
Each of the hyperuniform ensembles considered here is class III hyperuniform.
We note that $\bar{B}_{_V}s$ is minimized at $\beta = 1.3$, and tends to increase as $|\beta-1.3|$ increases, indicating that long-range translational order in strictly jammed binary disk packings tends to decrease at $\beta$ moves away from the center of the MRJ-like size ratio regime, which is consistent with $\alpha$ decreasing as $\beta$ moves away from the center of the MRJ-like size ratio regime. 

\begin{figure}
    \centering
    \includegraphics[width=0.45\textwidth]{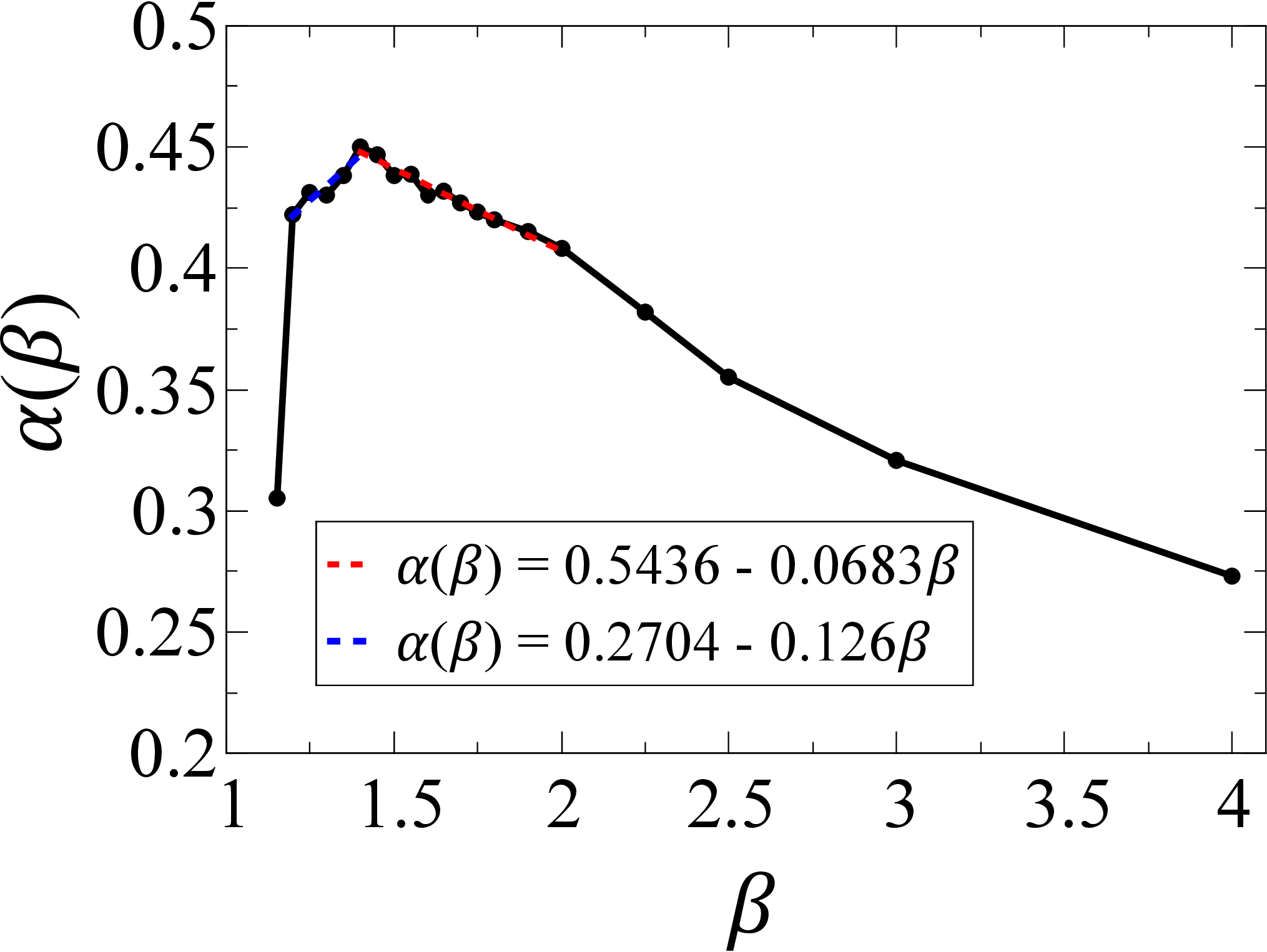}
    \caption{The hyperuniformity exponents $\alpha(\beta)$ extracted from the volume fraction variance curves as a function of the size ratio $\beta$ for the hyperuniform strictly jammed binary disk packing ensembles The black line corresponds to the extracted values and the blue and red dashed lines correspond to two linear fits of the data. 
    }
    \label{fig:avb}
\end{figure}

\begin{table}[]
\caption{The packing fraction $\phi$, rattler fraction $\phi_R$, percent of configurations that are isostatic in a given fixed-$\beta$ ensemble $\%_{Iso}$, the hyperuniformity scaling exponent $\alpha(\beta)$, the specific surface $s$ [cf. Eq. (\ref{eq:BinSS})], and the surface-area volume fraction variance scaling exponent $\bar{B}_{_V}$ scaled by the specific surface $s$ for each binary disk packing ensemble with size ratio $\beta$. Values of $\alpha$ and $B_{_V}s$ left blank correspond to nonhyperuniform ensembles.}

\label{tab:bin}
\begin{tabular}{lllllll}
\hline\hline
$\beta$    & $\phi$ & $\phi_R$ (\%) & $\%_{Iso}$& $\alpha(\beta)$ & $s$ &$\bar{B}_{_V}s\times10^{-3}$                    \\
\hline
1.01 & 0.8665 & 4.25          & 34.6     &                 & 1.724&                \\
1.05 & 0.8582 & 4.37          & 31.1     &                 & 1.674&                \\
1.1  & 0.8471 & 4.23          & 31.3     &                & 1.609&                \\
1.15 & 0.8406 & 4.06          & 30.1     & 0.306$\pm$0.008  & 1.556   & 6.90$\pm$0.03         \\
1.2  & 0.8384 & 3.99          & 35.2     & 0.422$\pm$0.004  & 1.512     & 6.12$\pm$0.02              \\
1.25 & 0.8384 &     4.05      &  38.2    &0.431$\pm$0.004   &  1.472    & 5.93$\pm$0.02              \\
1.3  & 0.8392 & 4.27          & 38.4     & 0.430$\pm$0.003  & 1.435     & 5.68$\pm$0.01              \\
1.35 & 0.8399 & 4.51          & 36.1     & 0.438$\pm$0.003  & 1.399    & 5.93$\pm$0.01              \\
1.4  & 0.8407 & 4.79          & 31.7     & 0.450$\pm$0.002  & 1.363     & 6.099$\pm$0.008             \\
1.45 & 0.8414 & 5.09          & 29.8     & 0.447$\pm$0.002  & 1.329&   6.42$\pm$0.02             \\
1.5 & 0.8422 & 5.40          & 32.9     & 0.438$\pm$0.003     & 1.296& 6.72$\pm$0.02              \\
1.55 & 0.8430 & 5.69          & 24.5     & 0.439$\pm$0.003    & 1.264& 7.11$\pm$0.02              \\
1.6  & 0.8437 & 6.02          & 27.3     & 0.430$\pm$0.003  & 1.232     & 7.00$\pm$0.01              \\
1.65 & 0.8444 & 6.31          & 25.0     & 0.432$\pm$0.003   & 1.202& 7.10$\pm$ 0.02             \\
1.7 & 0.8450 & 6.62          & 26.9     & 0.427$\pm$0.003       & 1.173& 6.99$\pm$0.02               \\
1.75 & 0.8456 & 7.00          & 21.3     & 0.423$\pm$0.003      & 1.145& 7.13$\pm$0.02              \\
1.8  & 0.8459 & 7.35          & 20.5     & 0.420$\pm$0.002  & 1.117      & 7.18$\pm$0.01              \\
1.9 & 0.8469 & 7.84          & 8.5     & 0.415$\pm$0.003       & 1.066& 7.03$\pm$0.02               \\
2.0  & 0.8478 & 8.12          & 9.8      & 0.408$\pm$0.002  & 1.017     & 6.70$\pm$0.01              \\
2.5  & 0.8512 & 14.21         & 0.4      & 0.355$\pm$0.004  & 0.822        & 8.31$\pm$0.02                  \\
3.0  & 0.8521 & 19.01         & 0.7      & 0.321$\pm$0.006  & 0.682        & 8.56$\pm$0.03                  \\ 
4.0 & 0.8557 & 26.11          & 0.0     & 0.273$\pm$0.006    & 0.503& 10.59$\pm$0.07             \\
5.0 & 0.8641 & 26.00          & 0.0     &                 & 0.399&                \\
6.0 & 0.8695 & 27.65          & 0.0     &                 & 0.329&                \\
7.0 & 0.8718 & 37.80          & 0.0     &                 & 0.279&                \\
8.0 & 0.8752 & 36.96          & 0.0     &                 & 0.242&                \\
\hline\hline
\end{tabular}
\end{table}

\subsection{Spectral Densities}\label{sec:binsd}
\begin{figure}
    \centering
    \includegraphics[width=0.4\textwidth]{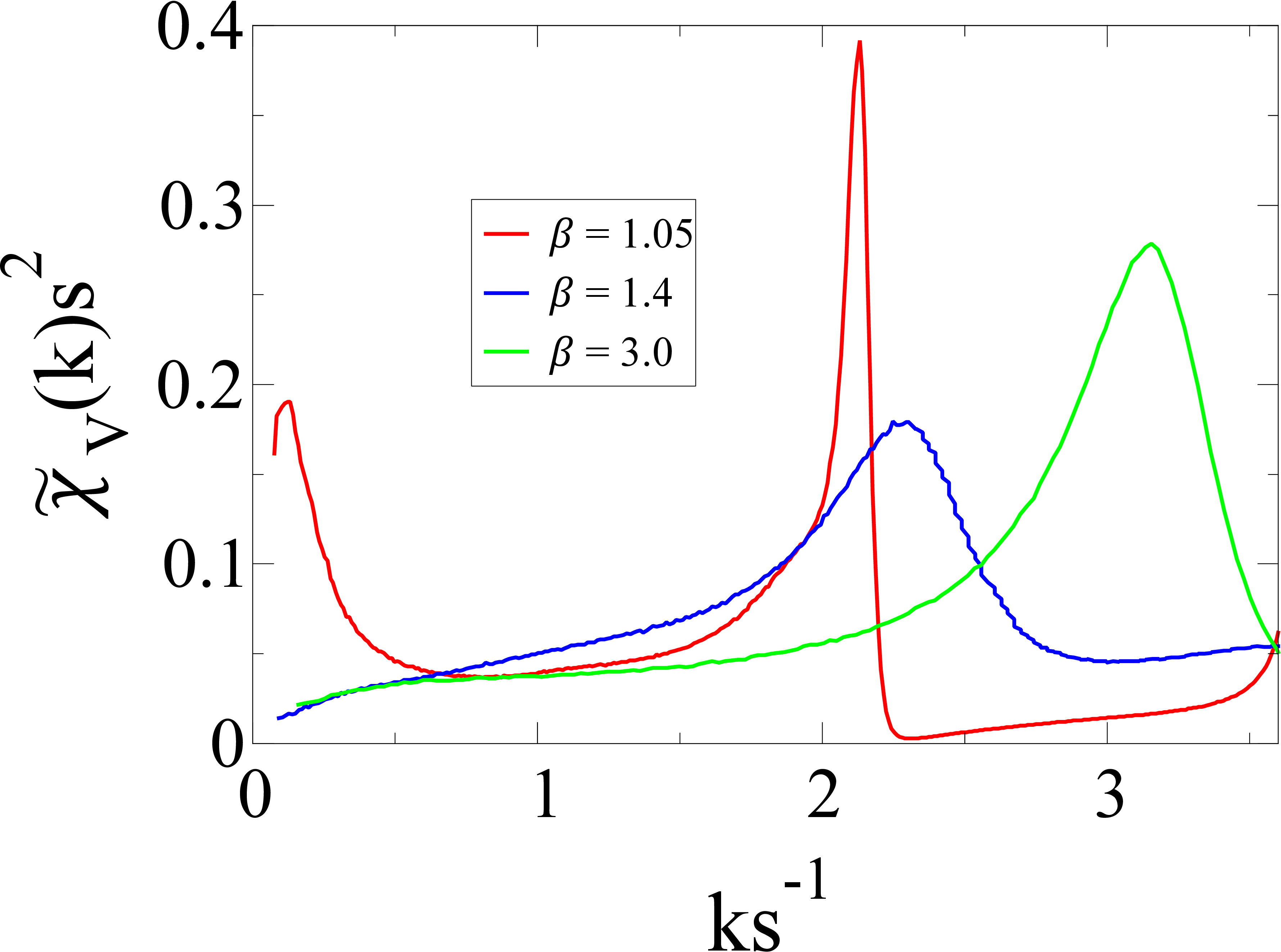}
    \caption{Ensemble averaged spectral densities scaled by the specific surface squared $\Tilde{\chi}_{_V}(k)s^2$ as a function of wavenumber scaled by the specific surface $ks^{-1}$ for select size ratios $\beta$ inside ($\beta=1.4$) and outside ($\beta=1.05$ and 3.0) the MRJ-like size ratio regime. 
    }
    \label{fig:bixk}
\end{figure}

In Fig. \ref{fig:bixk}, we show representative $\Tilde{\chi}_{_V}(k)$ curves for strictly jammed binary sphere packings with $\beta = 1.05, 1.4$, and 3.0.
Noise is present in these curves due to statistical fluctuations between different configurations in the same ensemble.
Each of the spectral density curves is consistent with disordered two-phase media, i.e., they disordered because they lack Bragg peaks.
The wavenumber at which the principle peak occurs corresponds to the length scale at which there are the greatest volume-fraction fluctuations.
Figure \ref{fig:bixk} shows that this length scale increases as $\beta$ decreases.
Moreover, the principle peaks tend to become more narrow as $\beta$ becomes small, indicating that the length scales at which these volume-fraction fluctuations occur have a smaller variance.
At intermediate length scales, small $\beta$ systems have lower scattering intensity, which can be attributed to the significant local ordering of the many crystallites in those packing ensembles, which are not present in as high a concentration as $\beta$ increases.
Finally, at large length scales (small-$k$), we can see the behavior captured by the $\sigma_{_V}^2(R)$ results above captured here.
Specifically, packings with $\beta$ in the MRJ-like regime have the lowest scattering intensity at small $k$, i.e., smaller large-wavelength volume-fraction fluctuations (which is consistent with larger $\alpha$, smaller $\bar{B}_{_V}$) and the scattering intensity at small $k$ tends to increase for packings with $\beta$ outside the MRJ-like regime.

\section{Spreadability}\label{sec:binspr}

\begin{figure}[b]
    \centering
    \includegraphics[width=0.4\textwidth]{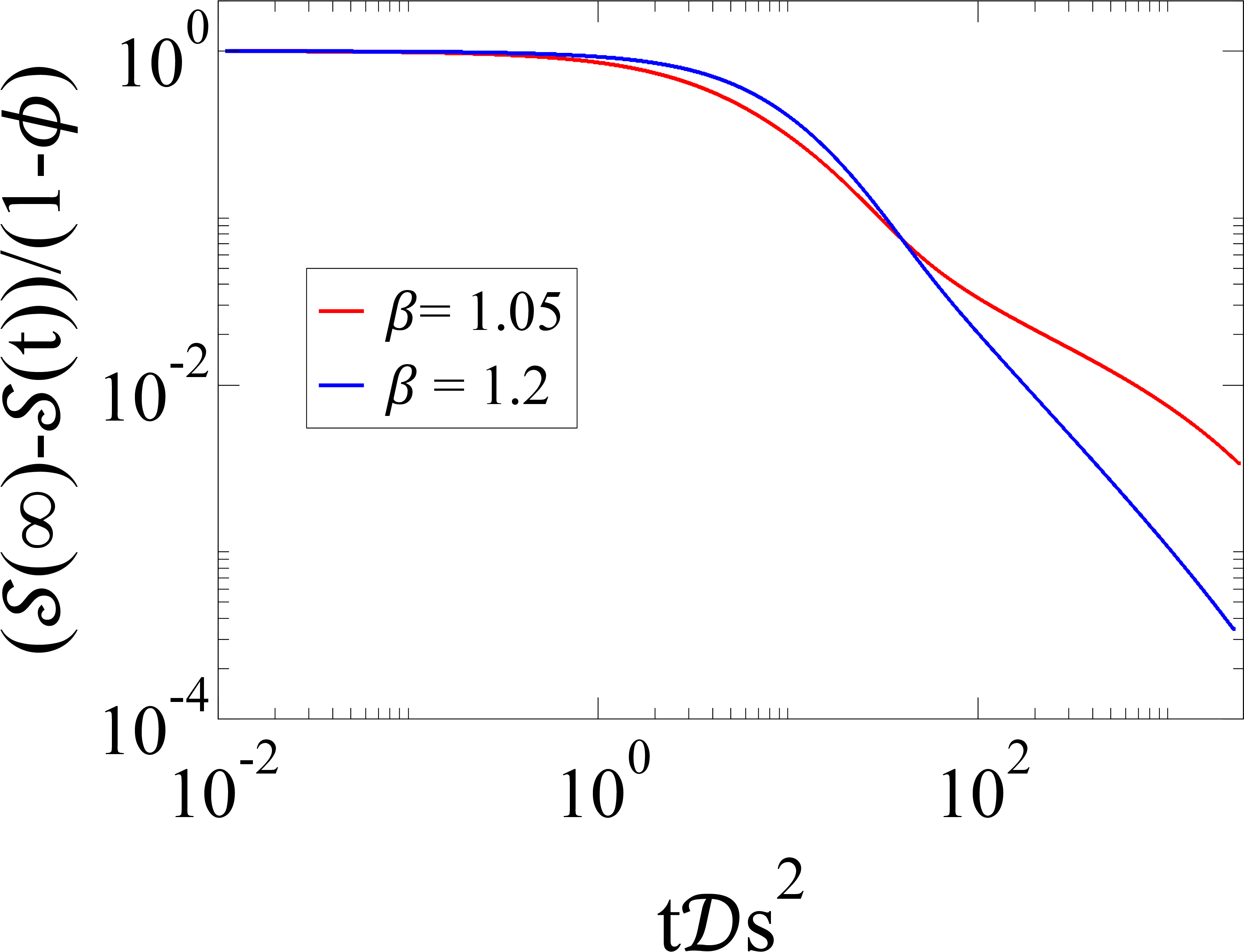}
    \caption{Representative curves of the excess spreadability $\mathcal{S}(\infty)-\mathcal{S}(t)$ scaled by $1-\phi$ as a function of dimensionless time $t\mathcal{D}s^2$ where $\mathcal{D}$ is the diffusion coefficient and $s$ is the specific surface. 
    }
    \label{fig:bin_spr}
\end{figure}

The spectral densities of the binary packings reported in the previous section can be utilized to estimate important effective physical properties of such media, including the effective dynamic dielectric constant \cite{To21_3,Vy23}, dynamic effective elastic constant \cite{Ki20}, fluid permeability \cite{To20}, trapping constant \cite{To20}, and diffusion spreadability \cite{To21_2}.
For the MRJ-like packings reported here, we compute the diffusion spreadability, which is exactly determined by the spectral density \cite{To21_2}.
Specifically, by the application of Eq. (\ref{eq:BiFT_spread}) together with the spectral densities reported in Sec. \ref{sec:BiRes}, we calculate the excess spreadability for each of the constant-$\beta$ ensembles produced in this work. 

In Fig. \ref{fig:bin_spr}, we show two representative examples of excess spreadability curves for our strictly jammed binary disk packings.
We note that, for the sake of visual clarity, we only present two excess spreadability curves, one far from the MRJ-like size ratio regime ($\beta = 1.05$), and one within said regime $\beta=1.2$.
Each of the excess spreadability curves for ensembles of packings from the family of MRJ-like packings behave similarly to the $\beta=1.2$ curve. 
The short-time behavior of the spreadability is determined by the specific surface $s$ of the packings given by Eq. (\ref{eq:BinSS}), specifically, to leading order in $t$ \cite{To21_2}
\begin{equation}
    \mathcal{S}(t) = \frac{s}{\phi}\left(\frac{\mathcal{D}t}{\pi}\right)^{1/2} + \mathcal{O}(\mathcal{D}t)\quad(t\rightarrow0),
\end{equation}
thus, media with larger $s$ will have faster short-time decay in their excess spreadabilities.
Indeed, the short-$t$ decay rates in Fig. \ref{fig:bin_spr} are consistent with the values of $s$ given in Table \ref{tab:bin}, which decrease with $\beta$.
At intermediate $t$, the small-$\beta$ packings begin to spread diffusion information more slowly than the MRJ-like packings, suggesting packings in the MRJ-like family have greater translational order on intermediate length scales.
At large times, the deviations between the two curves in Fig. \ref{fig:bin_spr} become more appreciable compared to those at intermediate times i.e., the gap between the two curves in Fig. \ref{fig:bin_spr} increases in magnitude. 
This behavior of the long-time excess spreadability is consistent with the hyperuniformity scaling exponents \textcolor{black}{of packings in the MRJ-like family} being larger than $\alpha$ for the disordered jammed packings outside of the family of MRJ-like packings.

\section{Conclusions}\label{sec:BiCon}

In this work, we used the TJ algorithm to generate a family of strictly jammed packings of equimolar binary hard disks with size ratios $\beta\in[1.01,8.0]$.
Using several criteria, specifically, the rattler fraction $\phi_R$, sixfold orientational order metric $\psi_6$, and the fraction of packings in a constant-$\beta$ ensemble that are isostatic, we show that the family of packings with $1.2\lesssim\beta\lesssim2.0$ are the most MRJ-like packings within the parameter space examined in this work.
Using the volume fraction variance $\sigma_{_V}^2(R)$, we find \textcolor{black}{that the hyperuniform packing ensembles are all of class III and the function $\alpha(\beta)$ is maximized at 0.450$\pm0.002$ when $\beta=1.4$ within the MRJ-like size ratio regime ($1.2\leq\beta\leq2.0$). Outside of this regime $\alpha(\beta)$ decays rapidly and for large $\beta$ or $\beta$ close to 1 the packings are nonhyperuniform.}
\textcolor{black}{Note that the well-studied case of the $\beta = 1.4$ equimolar disk system happens to produce the maximum $\alpha$ value.}
We also computed the window-surface-area coefficient $\bar{B}_{_V}$, and found that $\bar{B}_{_V}s$ tends to increase (i.e., the media have greater large-scale translational disorder) as $\beta$ deviated from the center of the MRJ-like size ratio regime, which is consistent with the decrease in $\alpha$.
Recall that we have used $\sigma_{_V}^2(R)$ to obtain $\alpha$ because, for finite nearly hyperuniform systems, variance-based methods are known to have a well-defined hyperuniform scaling regime that can be fit to precisely extract $\alpha$ \cite{To21}, which is not necessarily the case for $\Tilde{\chi}_{_V}(k)$ or spreadability-based fitting \cite{To21_2,Wa22, Ma22_2,Sk23,Ma23,Hi24} methods.
\textcolor{black}{When subsequently using $\Tilde{\chi}_{_V}(k)$ to compute the excess spreadability of these packings, we found that the packings in the MRJ-like family have similar long-time behaviors, which are distinct from the long-time behaviors of packings outside of the MRJ-like family.
In addition, we have computed the spectral density $\Tilde{\chi}_{_V}(k)$ of these packings, which clearly indicate they have scattering behavior consistent with disorder, i.e., they are disordered because they lack Bragg peaks.}



Our present study motivates the need for a more systematic investigation of what other disk-size distributions (both discrete and continuous) result in MRJ-like states and what the scaling laws are for MRJ-like packings in other parts of the disk-size distribution parameter space.
For example, when moving away from the 1:1 molar ratio one should not expect much change in $\alpha$ when $\beta$ is sufficiently small because the resulting size distribution would still be close to monodisperse regardless of the molar composition.
However, as $\beta$ increases, deviations from the 1:1 molar composition should begin to result in more rattlers if the large disk composition increases or a less uniform void space if the small disk composition increases, both of which are expected to decrease $\alpha$ \cite{To21, At16,Za11, Za11_2, Za11_3}. These effects should result in a greater reduction in $\alpha$ as $\beta$ increases.
By using a larger number of disk sizes (e.g., a ternary packing, or a packing with a continuous size distribution), the void spaces between disks should have a more uniform size distribution, which Refs. \cite{Za11, Za11_2, Za11_3} argue should increase $\alpha$.
It will be of great interest to consider conducting an analogous investigation of 3D MRJ-like packings of spheres with size-distributions to determine their corresponding scaling laws for such disordered jammed packings. In a previous study on disordered jammed binary sphere packings \cite{Ho13}, Hopkins \textit{et al.} report $\phi$ and $\phi_R$ for large region of the size ratio-molar composition parameter space. The aforementioned investigation could begin where $\phi_R$ is minimized in their parameter space.

As noted in Sec. \ref{sec:binspr}, the spectral density $\Tilde{\chi}_{_V}(\mathbf{k})$ (which we computed in Sec. \ref{sec:binsd}) can be used to estimate a number of other important effective physical properties of two-phase media, including the effective dynamic dielectric constant \cite{To21_3,Vy23}, dynamic effective elastic constant \cite{Ki20}, fluid permeability \cite{To20}, and trapping constant \cite{To20}.
As such, in future work it would be fruitful to more completely characterize how the effective physical properties of two-phase media derived from the packings examined herein change as a function of the disk size distribution.
The results from such a study, along with the spreadability results presented here, can be used to inform the design of hyperuniform thin-film materials using, e.g., Langmuir-Blodgett trough methods \cite{Da94, Ba06}, with tunable densities, degrees of orientational and translational order, and $\alpha$.

~
\begin{acknowledgments}
The authors thank H. Wang, M. Skolnick, P. K. Morse, and J. Kim for insightful discussions and valuable feedback on the manuscript. This research was sponsored by the Army Research Office and was accomplished under Cooperative Agreement No. W911NF22-2-0103 as well as the National Science Foundation under Award No. CBET-2133179.
\end{acknowledgments}




\begin{thebibliography}{93}
\bibitem [{Foo()}]{Foot4}%
  \BibitemOpen
  \href@noop {} {}\bibinfo {note} {A collection of recent papers that extract $\alpha$ from random close packed (RCP) \cite{Wi21,Wi23_2} and MRJ hypersphere packings \cite{Ma23} demonstrate the importance of using methods to extract $\alpha$ that can yield non-integer values of $\alpha$, unlike the polynomial fits used in, e.g., Refs. \cite{Do05, Za11, Za11_2}.}\BibitemShut {Stop}%
\end{thebibliography}

\providecommand{\noopsort}[1]{}\providecommand{\singleletter}[1]{#1}

\end{document}